\documentclass[prb]{revtex4}
\usepackage{times}
\usepackage{graphicx}
\usepackage{psfrag} 
\usepackage{fancyhdr}
\begin{document}
\date{\today}
\title{Electromagnetic braking: a simple quantitative model}
\author{\bf Yan Levin,  Fernando L. da Silveira, and  Felipe B. Rizzato} 
\affiliation{\it Instituto de F\'{\i}sica, Universidade Federal
do Rio Grande do Sul\\ Caixa Postal 15051, CEP 91501-970, 
Porto Alegre, RS, Brazil\\ 
{\small levin@if.ufrgs.br}}

\begin{abstract}

A calculation is presented which quantitatively accounts
for the terminal velocity of a cylindrical 
magnet falling through a long copper or aluminum pipe. The
experiment and the theory are a dramatic illustration of the Faraday's
and Lenz's laws and are bound to capture student's attention 
in any electricity and magnetism course.

\end{abstract}
\maketitle
\bigskip

\section{Introduction}

Take a long metal pipe made of a non-ferromagnetic material such as 
copper or aluminum, hold it vertically with respect
to the ground and place a small magnet at its top aperture.  The 
question is: when the magnet is released will it fall faster, slower,
or at the same rate as a non magnetic object of the same mass and shape? 
The answer is a dramatic demonstration of the Lenz's law, which never ceases
to amaze students and professors alike.  The magnet takes much
more time to reach the ground than a non-magnetic object. In fact we find
that for a copper pipe of length $L=1.7$m, the magnet takes more than
$20$s to fall to the ground, while a non-magnetic object covers 
the same distance in less than a second! Furthermore, when various magnets
are stuck together and then dropped through the pipe, the 
time of passage varies non-monotonically with the
number of magnets in the chain.  
This is contrary to the prediction of the point dipole
approximation which is commonly used to explain the 
slowness of the falling magnets~\cite{Sa92,MaBaBo93}.   
The easy availability of
powerful rare earth magnets, which can now be purchased in any toy store,
make this demonstration a ``must'' in any electricity and 
magnetism course~\cite{Sa92,MaBaBo93,HaJo98,PaCeHu05}.

In this paper we will go beyond a qualitative discussion of the dynamics
of the falling magnetic and present a theory which quantitatively
accounts for all the experimental findings.  The theory is
sufficiently simple that it should be easily accessible to 
students with only an intermediate exposure to the Maxwell's equations
in their integral form.

\section{Theory}

Consider a long vertical copper pipe of internal 
radius $a$ and wall thickness $w$.  
A cylindrical magnet of cross-sectional radius $r$, height $d$, and mass $m$ 
is held over its top aperture, see figure 1.
\begin{figure}[th]
\begin{center}
\includegraphics[width=5cm]{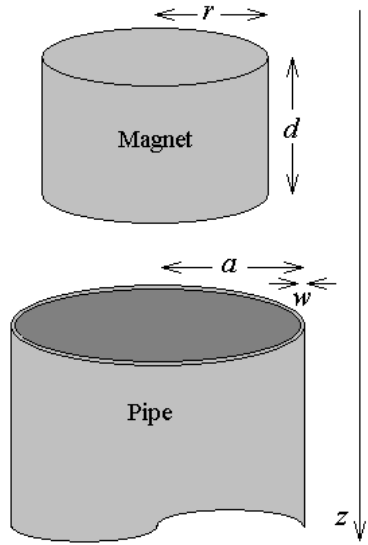}
\caption{The magnet and the pipe used in the experiment}
\end{center}
\end{figure}
It is convenient to imagine that the pipe is uniformly subdivided 
into parallel rings of width $l$. When the magnet is released,
the magnetic flux in each one of the rings begins to change. This, in 
accordance with the Faraday's law, induces an electromotive force 
and an electric current inside the ring. The magnitude
of the current will depend on the distance of each ring from the falling
magnet as well as on the magnet's speed. On the other hand, 
the law of Biot-Savart states that 
an electric current produces its own magnetic field which, 
according to the Lenz's law, must oppose the action that induced it
in the first place i.e. the motion of the magnet.  Thus, if the magnet is
moving away from a given ring the induced field will try to attract it 
back,
while if it is moving towards a ring the induced field will tend to repel
it. The net force on the magnet can be calculated by summing 
the magnetic interaction 
with  all the rings.  
The  electromagnetic force is an increasing function of the velocity and will
decelerate the falling magnet.  When the fall velocity reaches
the value at which the magnetic force completely compensates  
gravity, acceleration will go to zero and the magnet will continue
falling at a constant terminal velocity $v$. 
For a sufficiently strong magnet, the terminal velocity is reached very
quickly.  

It is interesting to consider the motion of the magnet from the
point of view of energy conservation.  When an object free
falls in a gravitational field, its potential energy is converted into kinetic
energy.  In the case of a falling magnet inside a copper pipe,
the situation is quite different.  Since the magnet 
moves at a constant velocity,
its kinetic energy does not change and  the gravitational potential
energy must be transformed into something else.  This ``something else''
is the ohmic heating of the copper pipe.  The gravitational energy must, 
therefore, be dissipated by the eddy currents 
induced inside the pipe.  In the steady
state the rate at which the magnet looses its gravitational energy 
is  equal to the rate of the energy dissipation by the ohmic resistance, 
\begin{equation}
\label{1}
m g v= \sum_z I(z)^2 R \;.
\end{equation}
In the above equation $z$ is the coordinate along the pipe length,  $I(z)$
is the current induced in the ring located at 
some position $z$, and $R$ is the resistance
of the ring.  
Since the time scales associated with the  
speed of the falling magnet are much larger than the ones associated
with the decay of eddy currents~\cite{Sm50,Sa92}, 
almost all the variation in electric current 
through a given ring results from the changing flux due to 
magnet's motion. The self-induction effects can thus be safely ignored.
Our goal, now, is to calculate the distribution of
currents in each ring, $I(z)$.  To achieve this we first study
the rate of change of the magnetic flux through one ring as the magnet
moves through the pipe. Before proceeding, however, we must first
address the question of the functional form of the magnetic field
produced by a stationary magnet. 
Since the magnetic permeability of copper and aluminum is very close
to that of vacuum, the magnetic field inside the pipe is 
practically identical to the one produced by the 
same magnet in vacuum. 
Normally this field
is approximated by that of a point dipole. This approximation is sufficient
as long as one wants to study the far field properties of the magnetic
field.  For a magnet confined to a pipe whose radius is comparable
to its size, this approximation is  no longer  valid.  
Since a large portion of the energy dissipation occurs 
in the near field, one would have 
to resum all of the  magnetic moments to
correctly account for the field in the magnet's vicinity.  
Clearly this is more work than can be done in a classroom demonstration.
We shall, therefore, take a different road.  Let us suppose that the
magnet has a uniform magnetization ${\bf M}=M {\bf \hat z}$. In this case the
magnetic charge density inside the magnet is zero,  while on the top
and the bottom of the magnet there is a uniform magnetic 
surface charge density $\sigma_M=M$ and $-\sigma_M=-M$ respectively.  The flux
produced by a cylindrical magnet can, therefore, be approximated 
by a field of two disks, each of radius $r$ separated by a distance
$d$.  Even this, however, is not an easy calculation, since the magnetic
field of a charged disk is a complicated 
quadrature involving Bessel functions.
We shall, therefore, make a further approximation and replace the
charged disks by point monopoles of the same net charge 
$q_m=\pi r^2 \sigma_M$.  The flux through a ring produced
by the two monopoles can now be easily calculated
\begin{equation}
\label{2}
\Phi(z)=\frac{\mu_0 q_m}{2} \left[\frac{z+d}{\sqrt{(z+d)^2+a^2}}-
\frac{z}{\sqrt{z^2+a^2}}\right]\;,
\end{equation}
where $\mu_0$ is the permeability of vacuum and 
$z$ is the distance from the nearest monopole, which we take to be 
the positively charged one, to the center
of the ring.   As the magnet falls, the flux through the ring changes, 
which results in an electromotive force given by the Faraday's law,
\begin{equation}
\label{3}
{\cal E}(z)=- \frac{{\rm d}\Phi(z) }{{\rm d} t}\;
\end{equation}
and an electric current
\begin{equation}
\label{4}
I(z)=\frac{\mu_0 q_m a^2 v}{2 R}\left[\frac{1}{(z^2+a^2)^{3/2}}-
\frac{1}{[(z+d)^2+a^2]^{3/2}}\right]\;.
\end{equation}
The rate of ohmic dissipation can now be calculated by evaluating the 
sum on the right hand side of Eq.~(\ref{1}). 
Passing to the continuum limit, we find the power dissipated to be
\begin{equation}
\label{5}
P=\frac{\mu_0^2 q_m^2 a^4 v^2}{4 R} \int_{-\infty}^\infty\frac{{\rm d} z}{l}
\left[\frac{1}{(z^2+a^2)^{3/2}}-
\frac{1}{[(z+d)^2+a^2]^{3/2}}\right]^2\;.
\end{equation}
Since most of the energy dissipation takes place near the magnet, 
we have explicitly extended the limits of integration to
infinity.  The resistance of each ring is $R= 2\pi a \rho/(w l)$, where $\rho$ 
is the electrical resistivity.  Eq.~(\ref{5}) can now
be rewritten as 
\begin{equation}
\label{6}
P=\frac{\mu_0^2 q_m^2 v^2 w}{8 \pi \rho a^2} f\left(\frac{d}{a}\right)\;,
\end{equation}
where $f(x)$ is a scaling function defined as
\begin{equation}
\label{7}
f(x)=\int_{-\infty}^\infty {\rm d} y
\left[\frac{1}{(y^2+1)^{3/2}}-
\frac{1}{[(y+x)^2+1]^{3/2}}\right]^2\;.
\end{equation}
Substituting Eq.~(\ref{6}) into  Eq.~(\ref{1}), 
the terminal velocity of a falling magnet is found to be
\begin{equation}
\label{8}
v=\frac{8 \pi m g \rho a^2}{\mu_0^2 q_m^2 w f\left(\frac{d}{a}\right)}\;.
\end{equation}
In figure 2 we plot the scaling function $f(x)$.
\begin{figure}[th]
\begin{center}
\includegraphics[width=8cm, angle=270]{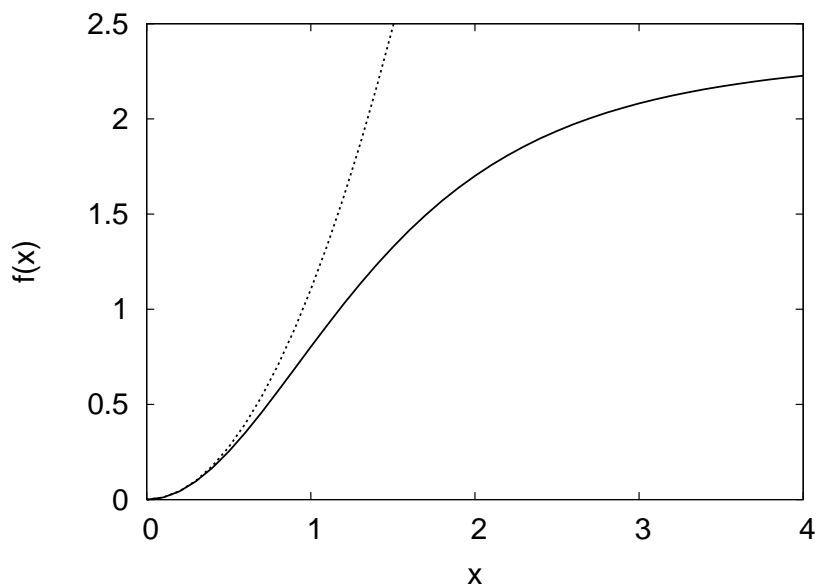}
\caption{The scaling function $f(x)$ (solid curve) and the limiting
form, Eq.~(\ref{9}) (dotted curve). Note the strong deviation
from the parabola (point dipole approximation) when $x>1$.}
\end{center}
\end{figure}
For small $x$  
\begin{equation}
\label{9}
f(x) \approx \frac{45 \pi}{128} x^2\;,
\end{equation}
and the terminal velocity reduces to~\cite{Sa92,MaBaBo93} 
\begin{equation}
\label{10}
v=\frac{1024}{45}\frac{m g \rho a^4}{\mu_0^2 p^2 w}\;,
\end{equation}
where $p=q_m d$ is the dipole moment of the falling magnet.  
We see, however, that as soon as the length of the magnet becomes
comparable to the radius of the pipe, the point dipole approximation
fails. In fact, for a realistic cylindrical magnets 
used in most demonstrations, one is always outside the
the point dipole approximation limit,
and the full expression (\ref{8}) must be used.

\section{Demonstration and discussion}

In our demonstrations we use a copper pipe (conductivity 
$\rho=1.75\times 10^{-8} \Omega$m)\cite{KoCh86}  of length $L=1.7$m, 
radius $a=7.85$mm, and wall thickness $w=1.9$mm;
three neodymium cylindrical magnets of mass $6$g each, radius
$r=6.35$mm, and height $d=6.35$mm; a stop watch; and a teslameter. 
We start by dropping one
magnet into the pipe and measure its time of passage --- $T=22.9$s.
For two magnets stuck together the time of passage increases to $T=26.7$s.
Note that if the point dipole approximation would be valid, the time
of passage would increase by a factor of two, which is clearly
not the case (within point dipole approximation the time of passage is directly
proportional to $p^2$ and inversely proportional to the mass, Eq.~(\ref{10}),
sticking two magnets together increases both the dipole moment 
and the mass of the magnet by a factor of two).  
Furthermore, when all three magnets are stuck together
the time of passage drops to $T=23.7$s.  Since the 
terminal velocity is reached
very quickly, a constant speed of 
fall approximation is justified for the whole
length of the pipe.  
In Table 1 we present the values for the 
measured velocity $v=L/T$. We next compare 
this measurements with the predictions of the theory. 
First, however, we have to obtain the value of $q_m$ for
the  magnet.  
To do this we measure the magnetic field at 
the {\it center} of one of the {\it flat surfaces} 
of the magnet using the digital teslameter {\it Phywe} 
(the probe of teslameter
is brought in direct contact with the surface of the magnet).  
Within our uniform magnetization
approximation this field is produced by two parallel disks of radius $r$ 
and magnetic surface charge $\pm \sigma_M$, separated by distance $d$,
\begin{equation}
\label{11}
B=\frac{\mu_0 \sigma_m}{2} \frac{d}{\sqrt{d^2+r^2}}\;.
\end{equation}
The  magnetic charge is, therefore,
\begin{equation}
\label{12}
q_m=\frac{2 \pi B r^2 \sqrt{d^2+r^2}}{\mu_0 d}\;.
\end{equation}
For $n$ magnets associated in series $d \rightarrow h=n d$ and
one can check using the values of the measured magnetic field
presented in the Table 1
that $q_m$ is invariant of $n$, up to experimental error, 
justifying our uniform magnetization
approximation. Rewriting Eq.~(\ref{8}) in terms of the measured
magnetic field for a combination of $n$ magnets, we arrive at
\begin{equation}
\label{13}
v=\frac{2 M g \rho a^2 h^2}{\pi B^2 r^4 w (h^2+r^2) f\left(\frac{h}{a}\right)}\;,
\end{equation}
where $M=n m$ and $h=n d$.  In Table 1 we compare the values of the
measured and the calculated terminal velocities.
\begin{table}[h]
\caption{Experimental and theoretical 
values of the terminal velocity }
\vspace{.2cm}
\centering
\begin{tabular}{|c|c|c|c|} \hline
$n$ magnets& $B$(mT) & $v_{exp}$ ($10^{-2}$ m/s) & $v_{theory}$ ($10^{-2}$ m/s) \\
\hline \hline \hline \hline
1 & 393 & 7.4 & 7.3\\ \hline
2 & 501 & 6.4 & 5.8   \\ \hline
3 & 516 & 7.2 & 6.9  \\ \hline
\end{tabular}
\end{table}
   
Considering the complexity of the problem, the simple theory presented
above accounts quite well for all the experimental findings.  In particular,
the theory correctly predicts that two magnets stuck together fall slower
than either one magnet separately or all three magnets together.  
For each pipe there is, therefore, an optimum magnetic 
size which falls the slowest.

\section{Conclusions}

We have presented a simple theory which accounts for the
electromagnetic braking of a magnet falling through a conducting pipe.  
The experiment is a dramatic illustration of the Faraday's and 
Lenz's law. Perhaps surprisingly, a quantitative discussion of the
experiment is possible with only a basic 
knowledge of electrodynamics. 
Furthermore, the only specialized equipment necessary for performing
the measurements is a teslameter, which is usually present in any physics
laboratory.  The demonstration and the calculations presented in this paper
should, therefore, be easily adoptable to almost
any electricity and magnetism course.

\bibliographystyle{prsty}
\bibliography{references}

\begin{thebibliography}{1}

\bibitem{Sa92}
W.~M. Saslow, Am. J. Phys. {\bf 60},  693  (1992).

\bibitem{MaBaBo93}
C.~S. MacLatchy, P. Backman, and L. Bogan, Am. J. Phys. {\bf 61},  1096
  (1993).

\bibitem{HaJo98}
K.~D. Hahn, E.~M. Johnson, A. Brokken, and S. Baldwin, Am. J. Phys. {\bf 66},
  1066  (1998).

\bibitem{PaCeHu05}
J.~A. Palesko, M. Cesky, and S. Huertas, Am. J. Phys. {\bf 73},  37  (2005).

\bibitem{Sm50}
W.~R. Smythe, {\em Static and {D}ynamic {E}lectricity} (McGraw-Hill, New York,
  1950).

\bibitem{KoCh86}
N.~I. Kochkin and M.~G. Chirk\'evitch, {\em Prontu\'ario de f\'{\i}sica
  elementar} (MIR, Moscow, 1986).

\end{thebibliography}
\end{document}